\documentclass[twoside]{article}
\usepackage{fleqn,espcrc2}

\usepackage{graphicx,graphics}
\usepackage[figuresright]{rotating}

\newcommand{\rf}[1]{(\ref{#1})}
\newcommand{\beq}{\begin{equation}}
\newcommand{\eeq}{\end{equation}}
\newcommand{\bea}{\begin{eqnarray}}
\newcommand{\eea}{\end{eqnarray}}

\newcommand{\e}{\mbox{e}}

\renewcommand{\a}{\alpha}

\newcommand{\Del}{\Delta}

\newcommand{\ra}{\rangle}
\newcommand{\la}{\langle}

\newcommand{\mi}{\!-\!}
\newcommand{\equ}{\!=\!}
\newcommand{\pl}{\!+\!}

\newcommand{\AmS}{{\protect\the\textfont2
  A\kern-.1667em\lower.5ex\hbox{M}\kern-.125emS}}

\title{Computer Simulations of 3d Lorentzian Quantum Gravity}
\author{J. Ambj\o rn\address{The Niels Bohr Institute, 
Blegdamsvej 17, DK-2100 Copenhagen \O , Denmark\\}\thanks{ 
Supported by ``MaPhySto'', Centre of Mathematical Phy\-sics 
and Stochastics -- financed  by the 
National Danish Re\-search Foundation.}${}^{\ddag}$,
J. Jurkiewicz\address{Institute of Physics,
Jagellonian University,
Reymonta 4, PL 30-059 Krakow, Poland}${}^{*}$\thanks{Supported 
by KBN grants
2\,P03B 019\,17 and 008\,14.}${}^{\ddag}$
and R. Loll\address{Albert-Einstein-Institut,
Max-Planck-Institut f\"{u}r Gravitationsphysik,\\
Am M\"uhlenberg 1, D-14476 Golm, Germany}\thanks{
Supported by EU network on ``Discrete Random Geometry'', 
grant HPRN-CT-1999-00161.}}
       
\begin{document}

\begin{abstract}
We investigate the phase diagram of non-perturbative 
three-dimensional Lorentzian quantum gravity
with the help of Monte Carlo simulations.
The system has a first-order phase transition at a critical
value $k_0^c$ of the bare inverse gravitational coupling 
constant $k_0$. For $k_0 > k_0^c$ the system reduces to a 
product of uncorrelated Euclidean 2d gravity models and 
has no intrinsic interest as a model of 3d gravity. For $k_0 < k_0^c$,
extended three-dimensional geo\-metries dominate
the functional integral despite the 
fact that we perform a sum over geometries and no particular background 
is distinguished at the outset.
Furthermore, all systems with $k_0 < k_0^c$ have the same 
continuum limit. A different choice of $k_0$ corresponds merely 
to a redefinition of the overall length scale.
\vspace{1pc}
\end{abstract}

\maketitle

\section{INTRODUCTION}

We still need to find {\it the
theory of quantum gravity}.
In \cite{al,ajl} we have proposed a dynamically triangulated 
model of two, three and four dimensional 
quantum gravity with the following properties:\\
(a) Lorentzian space-time geometries (histories) are
obtained by causally gluing sets of Lorentzian building 
blocks, i.e. $d$-dimensional simplices with simple length
assignments;\\
(b) all histories have a preferred discrete notion of 
proper time $t$; $t$ counts the number of evolution 
steps of a transfer matrix between adjacent spatial
slices, the latter given by $(d\mi 1)$-dimensional 
triangulations of equilateral Euclidean simplices;\\
(c) each Lorentzian discrete geometry can be ``Wick-rotated"
to a Euclidean one, defined on the same (topological) 
triangulation; \\
(d) at the level of the discretized action, the ``Wick rotation"
is achieved by an analytic continuation in the dimensionless
ratio $\alpha\equ -l^{2}_{\rm t}/l^{2}_{\rm s}$ 
of the squared time- and space-like link length; 
for $\alpha\equ -1$ one obtains
the usual Euclidean action of dynamically triangulated gravity;\\
(e) the extreme phases of degenerate geometries found in
the Euclidean models cannot be realized in the Lorentzian
case.

A metric space-time is constructed by ``filling in"
for all $t$ the $d$-dimensional sandwich between 
the pair of spatial slices at integer times $t$ and $t\pl 1$.
We only consider regular gluings which lead to {\it simplicial
manifolds}. In the case of {\it three}-dimensional quantum gravity 
the basic building blocks are three types of 
Lorentzian tetrahedra, 
(1): (3,1)-tetrahedra (three vertices contained in slice $t$ and
one vertex in slice $t+1$): they have three space- and three time-like
edges; their number in the sandwich $[t,t\pl 1]$ 
will be denoted by $N_{31}(t)$; 
(2): (1,3)-tetrahedra: the same as above, but upside-down; the
tip of the tetrahedron is at $t$ and its base lies in the slice
$t+1$; notation $N_{13}(t)$;
(3): (2,2)-tetrahedra: one edge (and therefore two vertices)
at each $t$ and $t+1$; they have two space- and four time-like edges;
notation $N_{22}(t)$.

Each of these triangulated space-times carries a causal
structure defined by the piecewise flat Lorentzian metric. 
Each time-like link is given a future-orientation
in the positive $t$-direction, so that two lattice vertices connected by
a sequence of positively oriented links are causally
related.  

As mentioned in (d) above, after rotating to Euclidean 
signature and choosing $\a= \mi 1$, the Einstein action 
becomes identical to the action
\beq\label{euact2}
S_{\rm E}(N_0,N_3,T) = -k_0 N_0 +k_3 N_3
\eeq
used in Euclidean dynamical triangulations \cite{d4dt}, 
where $N_0$ and $N_3$ denote the total numbers of vertices and tetrahedra 
in the triangulation. 
The dimensionless couplings $k_0$ and $k_3$ have been introduced 
to conform with the conventions used in 3d {\it Euclidean}
simplicial quantum
gravity. Here $k_0$ is proportional to 
the bare inverse gravitational coupling constant, while $k_3$ is a 
linear combination of the bare cosmological and inverse 
gravitational constants. 
In a slight abuse of language, we will
still refer to $k_3$ as the (bare) cosmological constant.
We fix the space-time topology to $S^1\times S^2$, where the
periodic identification in the $t$-direction has been
chosen entirely for practical convenience. The total length
of the space-time in the compactified $S^1$-direction (i.e.
the number of proper-time steps) is denoted by $T$. 
The partition function is thus
\beq\label{wick2}
Z(k_0,k_3,T)=\sum_{{\cal T}_{T}(S^1\times S^2)} 
\e^{- 
S_{\rm E}(N_0,N_3,T)},
\eeq
where the sum is taken over the class ${\cal T}_{T}(S^1\times S^2)$
of triangulations of $S^1\times S^2$ specified earlier,
for fixed $T$.
While \rf{wick2} is similar in structure to the standard 
partition function of
dynamically triangulated Euclidean gravity,
one should bear in mind that the set of triangulations 
contributing in \rf{wick2} is quite different.

\section{THE MONTE CARLO SIMULATION}

We explore the phase diagram of the theory defined by the 
state sum \rf{wick2} using Monte Carlo simulations. 
A 3d triangulation in the sum \rf{wick2}
consists of $t$ successive 2d triangulations. 
These spatial slices are glued together by filling the
space-time gaps between them with the three types of building blocks 
described in the introduction.

A local updating algorithm consisting of five basic {\it moves}
changes one such triangulation 
into another one, while preserving the constant-time slice structure. 
A successive application of these moves will take us  
around in the class of triangulation with fixed $T$. 
The moves are (see
\cite{xxxx} for more details):\\
(1): consider two neighbouring triangles in the spatial $t$-plane
such that the two associated pairs of 
tetrahedra above and below the $t$-plane each share one triangle.
We can perform a standard ``flip''
move of the link common to the two triangles in the $t$-plane and 
make a corresponding reassignment of tetrahedra.
(2\&3): Consider a triangle in the spatial $t$-plane and
insert a vertex at its centre. In this way the 
two neighbouring tetrahedra sharing the triangle are replaced by 
six tetrahedra, three above 
and three below. There is an obvious inverse move.
(4\&5): The fourth move is the standard Alexander move performed 
on a (2,2)-tetrahedron and a (3,1)-tetrahedron sharing a triangle,  
replacing it by two (2,2)- and 
one (3,1)-tetrahedra, and the fifth move is its inverse. Obviously 
the (3,1)-tetrahedron could have been replaced by a (1,3)-tetrahedron 
in (4\&5). 

The strategy for the simulations is the usual one from dynamical 
triangulations: fine-tune the cosmological constant to its critical value
$k^c_3(k_0)$ (which depends on the bare inverse gravitational coupling 
constant $k_0$) and keep the fluctuations of space-time volume bounded
within a certain range. Then measure expectation values of suitable
observables for these quantum universes.

\section{RESULTS} 

\begin{figure}[h]
\vspace{-2.5cm}
\centerline{\scalebox{0.4}{\rotatebox{0}{\includegraphics{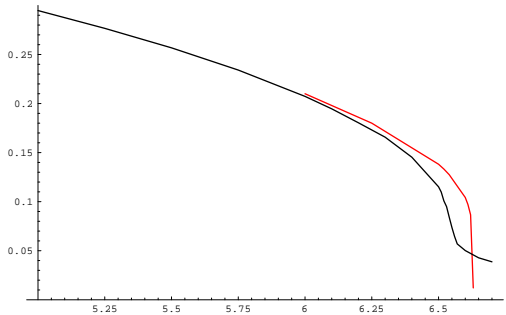}}}}
\vspace{-3.3cm}
\caption[fig6]{The order parameter $\tau = N_{22}/N_{tot}$ for configurations
with $T \equ 64$, and $N_3\equ 16,000$ (black curve) and 
64,000 (red curve), plotted as a function of $k_0$. }
\label{fig6}
\end{figure}

In order to explore the phase diagram of our regularized model 
we have to identify an order parameter, and explore how it 
changes with the coupling constant, in this case $k_0$.  
We have found that the ratio $\tau$ between 
the total number $N_{22}$ of (2,2)-tetrahedra 
and the total space-time volume 
$$
N_{3}\equiv N_{tot}= N_{22}+N_{31}+N_{13}
$$
serves as an efficient order parameter. 
In Fig.\ \ref{fig6}
we show $\tau$ as a function of $k_0$. One
observes a rapid drop to zero of $\tau(k_0)$ around $k_0 \approx 6.64$.
Increasing $N_3$, the drop becomes a jump, typical for a 
first-order phase transition. A detailed study of the neighbourhood 
of $k_0 \equ 6.64$ reveals a hysteresis as one performs a cycle, 
moving above 
and below the critical value $k_0^c$, again as expected from
a first-order transition.  

The phase above $k_0^c$ is of no interest for continuum 3d quantum gravity
since one can show that it is equivalent to an uncorrelated product of 
2d Euclidean gravity models \cite{xxxx}.

\begin{figure}[h]
\vspace{-2.0cm}
\centerline{\scalebox{0.4}{\rotatebox{0}{\includegraphics{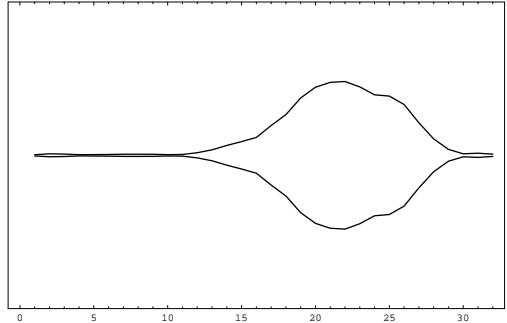}}}}
\vspace{-3.3cm}
\caption[fig10]{Snapshot of the distribution of 2d volumes 
$N_{31}(t)$ of spatial slices at times $t\in [0,T]$, with $T \equ 
32$ and $k_0 \equ 5.0$ ($k_0^c \equ 6.64$).}
\label{fig10}
\end{figure}

\begin{figure}[h]
\vspace{-2.0cm}
\centerline{\scalebox{0.4}{\rotatebox{0}{\includegraphics{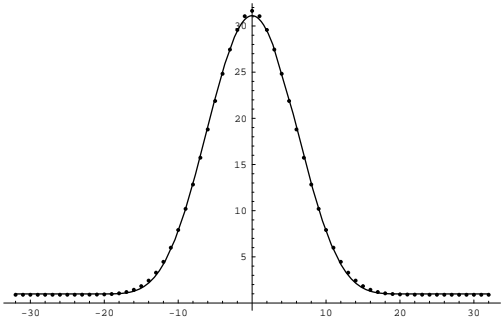}}}}
\vspace{-3cm}
\caption[fig11]{The correlator $C(\Del)$ with 
$T\equ 64$ and $N_{3}\equ 32,000$. Dots are the measured 
values (error bars less than dots), and the curve is fitted from 
the sphere solution described in the text.}
\label{fig11}
\end{figure}

We now turn to the interesting phase of $k_0 < k_0^c$. 
Remarkably, we observe here the 
emergence of well-defined three-dimensional configurations.
In Fig.\ \ref{fig10} we present a snapshot of a 
configuration of 16,000 
tetrahedra for $k_0\equ 5.0$ and of total proper-time extent $T\equ 32$. 
It shows the 2d spatial volume as a function of the time $t$.
Following the computer-time history 
of this extended object, it is clear that although it does indeed fluctuate,
the fluctuations take place around a three-dimensional object of a
well-defined linear extension. {\it The emergence of a ground state of
extended geometry is a very non-trivial property
of the model, since we never put in any preferred background geometry
by hand.} No structures of this kind have ever been observed in
dynamically triangulated models of {\it Euclidean} quantum gravity.
It underscores the fact that the Lorentzian models are genuinely
different and affirms our conjecture that in $d\geq 3$ they are
less pathological than their Euclidean counterparts.
Let us denote the typical time-extent of our extended ``universe'' by
$T_u$. We will always choose
the total proper time $T$ sufficiently large, 
such that $T_u < T$ for the range of $N_3$ under consideration.
Since both the shape of the universe and its location along the
$t$-direction fluctuate, we have found it convenient to measure
the correlation function 
\beq\label{n5}
C(\Del) = \frac{1}{{T}^2}\sum_{t=1}^{T}\la N_{2s}(t)N_{2s}(t+\Del)\ra,
\eeq 
where $N_{2s}(t)\equ N_{13}(t)+N_{31}(t)+N_{22}(t)$, 
as a function of the displacement $\Del$
to determine the scaling of $T_u$ with the
space-time volume $N_3$.
This correlator has the advantage of 
being translation-invariant in $t$ and allows for a precise 
measurement by averaging over many independent configurations. 
From the typical shape of the space-time configurations 
we expect $C(\Del)$ to be of the order of the spatial cut-off 
if $\Del \gg 2 T_b$. Fig.\ \ref{fig11} illustrates the result of  
our measurements of $C(\Del)$, with the dots representing the
measured values.
The theoretical curve to which we are fitting corresponds to a 
sphere $S^3$ with the radius as a free parameter. 
(By $S^3$ we mean the 3d geo\-metry of constant positive curvature
which is a {\it classical} solution of Euclidean gravity with a
positive cosmological constant.)
In order to ``adapt'' the $S^3$-solution to the topology $S^1\times S^2$,
we assume that the standard
$S^3$-solution is valid until the radius $r$ of the 
$S^2$-slice reaches the spatial cut-off scale. Beyond this point,
the value of $r$ is frozen to the cut-off value, which we also take 
as a free parameter. 
For this ``spherical'' geometry we then perform the integral (the sum) 
in \rf{n5}, without the average $\la \cdot \ra$.
As is evident from Fig.\ \ref{fig11}, the volume distribution
associated with this fixed geometry gives a rather good fit to
our data. This provides some evidence that we can 
ignore the quantum average implied by $\la \cdot \ra$, and that 
our universes behave
semi-classically, {\it at least
as far as their macroscopic geometric properties are concerned}.
We should mention that our ``$S^3$-solution'' is not
singled out uniquely, since the choice of a Gaussian shape in the 
$t$-direction gives a fit of comparable quality. 

For various space-time volumes $N_{3}$ (typically 8, 16, 32 and 64k)
we have determined the radius $R_{S^3}$ of $S^3$ 
from the fits to the measured 
$C(\Del)$. From this, we have finally found $\a = 0.34\pm 0.02$
as the best exponent in the scaling relation
\beq\label{n6}
R_{S^3}(N_{3}) = N_{3}^\a.
\eeq
The same value is obtained using other ways to extract $T_u$, 
lending additional support to the genuinely three-dimensional
nature of our universes.

\begin{figure}[h]
\vspace{-2.5cm}
\centerline{\scalebox{0.4}{\rotatebox{0}{\includegraphics{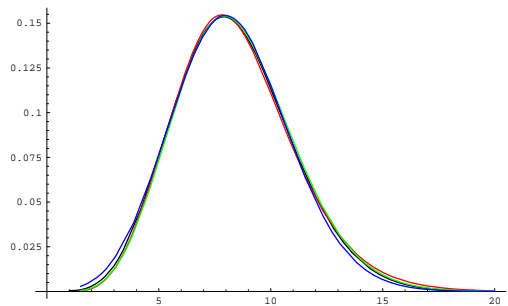}}}}
\vspace{-3cm}
\caption[fig13]{The 2d volume $V(l)$ of spatial spherical shells,
measured only on slices inside the spherical universe,
for various $k_0$ and rescaled, using 
the seemingly unrelated scaling obtained by applying the same 
philosophy to the correlator $C(\Del)$.}
\label{fig13}
\end{figure}

Another important result concerns 
the relation between the geometries of different $k_0$, in the phase
where $k_0 <k_0^c$. In the numerical simulations we have observed
the following: 
(i) the distributions as functions of 
$t$ can be made to coincide for different $k_0$ 
by rescaling the time, $t \to f_{ti}(k_0) t$ or 
$a_t \to f_{ti}(k_0) a_t$, where $a_t$ is the link length in time 
direction. (ii) The distributions measured in the spatial slices 
from {\it inside} the universe can be made to 
coincide for different $k_0$ by rescaling the 
spatial link distance $a_s \to f_{sp}( k_0) a_s$, where $a_s$ is 
the length of the spatial links. 
This is illustrated in Fig.\ \ref{fig13} for the distributions 
of 2d volumes $V(l)$ of {\it spatial} spherical shells of 
(link) radius $l$, measured for various values of $k_0$. 
(iii) Within the numerical accuracy we find that
$f_{ti}(k_0)=f_{sp}(k_0)$. 

We conclude that apart from the overall length scale of the universe, 
set by the bare inverse gravitational coupling $k_0$,
{\it the physics is the same for all values $k_0$ below $k_0^c$.}

\end{document}